\RequirePackage{lineno} 
\documentclass[aps,prl,twocolumn,showpacs,superscriptaddress]{revtex4}
\usepackage{graphicx}
\usepackage{dcolumn}
\usepackage{array}
\usepackage{bm}
\usepackage{color}
\usepackage{lineno}

\newcommand{\dzero}{\mbox{$D^0$}}

\newcommand{\sqrtsNN}{\mbox{$\sqrt{s_{_{\mathrm{NN}}}}$}}
\newcommand{\dAu}{\textit{d}+Au}

\newcommand{\AuAu}{Au+Au}

\renewcommand{\AA}{\mbox{A+A}}

\newcommand{\pp}{\mbox{\textit{p}+\textit{p}}}

\newcommand{\pt}{\mbox{$p_{\rm T}$}}

\newcommand{\gevcc}{\mbox{$\mathrm{GeV/}c^2$}}

\newcommand{\gevc}{\mbox{${\mathrm{GeV/}}c$}}

\newcommand{\RAA}{\mbox{$R_{\rm AA}$}}

\newcommand{\dedx}{\mbox{$dE/dx$}}
\newcommand{\npart}{\mbox{$N_{\mathrm{part}}$}}
\newcommand{\nbin}{\mbox{$N_{\mathrm{bin}}$}}
\newcommand{ \la }{\langle}
\newcommand{ \ra }{\rangle}

\newcolumntype{d}[1]{D{.}{\cdot}{#1}}

\begin{document}

\def\Journal#1#2#3#4{{#1} {\bf #2}, #3 (#4)}

\def\NCA{\em Nuovo Cimento}
\def\NIM{\em Nucl. Instr. Meth.}
\def\NIMA{{\em Nucl. Instr. Meth.} A}
\def\NPB{{\em Nucl. Phys.} B}
\def\NPA{{\em Nucl. Phys.} A}
\def\PLB{{\em Phys. Lett.} B}
\def\PRL{{\em Phys. Rev. Lett.}}
\def\PRC{{\em Phys. Rev.} C}
\def\PRD{{\em Phys. Rev.} D}
\def\ZPC{{\em Z. Phys.} C}
\def\JPG{{\em J. Phys.} G}
\def\EPJ{{\em Eur. Phys. J.} C}
\def\RPP{{\em Rep. Prog. Phys.}}
\def\IPA{{\em Int. J. Mod. Phys.} A}
\def\IPE{{\em Int. J. Mod. Phys.} E}
\def\JHEP{{\em J. High Energy Phys.}}

\title{Observation of $D^0$ meson nuclear modifications in Au+Au collisions at \sqrtsNN = 200\,GeV}

\date{\today}
\affiliation{AGH University of Science and Technology, Cracow, Poland}
\affiliation{Argonne National Laboratory, Argonne, Illinois 60439, USA}
\affiliation{University of Birmingham, Birmingham, United Kingdom}
\affiliation{Brookhaven National Laboratory, Upton, New York 11973, USA}
\affiliation{University of California, Berkeley, California 94720, USA}
\affiliation{University of California, Davis, California 95616, USA}
\affiliation{University of California, Los Angeles, California 90095, USA}
\affiliation{Universidade Estadual de Campinas, Sao Paulo, Brazil}
\affiliation{Central China Normal University (HZNU), Wuhan 430079, China}
\affiliation{University of Illinois at Chicago, Chicago, Illinois 60607, USA}
\affiliation{Cracow University of Technology, Cracow, Poland}
\affiliation{Creighton University, Omaha, Nebraska 68178, USA}
\affiliation{Czech Technical University in Prague, FNSPE, Prague, 115 19, Czech Republic}
\affiliation{Nuclear Physics Institute AS CR, 250 68 \v{R}e\v{z}/Prague, Czech Republic}
\affiliation{Frankfurt Institute for Advanced Studies FIAS, Germany}
\affiliation{Institute of Physics, Bhubaneswar 751005, India}
\affiliation{Indian Institute of Technology, Mumbai, India}
\affiliation{Indiana University, Bloomington, Indiana 47408, USA}
\affiliation{Alikhanov Institute for Theoretical and Experimental Physics, Moscow, Russia}
\affiliation{University of Jammu, Jammu 180001, India}
\affiliation{Joint Institute for Nuclear Research, Dubna, 141 980, Russia}
\affiliation{Kent State University, Kent, Ohio 44242, USA}
\affiliation{University of Kentucky, Lexington, Kentucky, 40506-0055, USA}
\affiliation{Korea Institute of Science and Technology Information, Daejeon, Korea}
\affiliation{Institute of Modern Physics, Lanzhou, China}
\affiliation{Lawrence Berkeley National Laboratory, Berkeley, California 94720, USA}
\affiliation{Massachusetts Institute of Technology, Cambridge, Massachusetts 02139-4307, USA}
\affiliation{Max-Planck-Institut f\"ur Physik, Munich, Germany}
\affiliation{Michigan State University, East Lansing, Michigan 48824, USA}
\affiliation{Moscow Engineering Physics Institute, Moscow Russia}
\affiliation{National Institute of Science Education and Research, Bhubaneswar 751005, India}
\affiliation{Ohio State University, Columbus, Ohio 43210, USA}
\affiliation{Old Dominion University, Norfolk, Virginia 23529, USA}
\affiliation{Institute of Nuclear Physics PAN, Cracow, Poland}
\affiliation{Panjab University, Chandigarh 160014, India}
\affiliation{Pennsylvania State University, University Park, Pennsylvania 16802, USA}
\affiliation{Institute of High Energy Physics, Protvino, Russia}
\affiliation{Purdue University, West Lafayette, Indiana 47907, USA}
\affiliation{Pusan National University, Pusan, Republic of Korea}
\affiliation{University of Rajasthan, Jaipur 302004, India}
\affiliation{Rice University, Houston, Texas 77251, USA}
\affiliation{University of Science and Technology of China, Hefei 230026, China}
\affiliation{Shandong University, Jinan, Shandong 250100, China}
\affiliation{Shanghai Institute of Applied Physics, Shanghai 201800, China}
\affiliation{SUBATECH, Nantes, France}
\affiliation{Temple University, Philadelphia, Pennsylvania 19122, USA}
\affiliation{Texas A\&M University, College Station, Texas 77843, USA}
\affiliation{University of Texas, Austin, Texas 78712, USA}
\affiliation{University of Houston, Houston, Texas 77204, USA}
\affiliation{Tsinghua University, Beijing 100084, China}
\affiliation{United States Naval Academy, Annapolis, Maryland, 21402, USA}
\affiliation{Valparaiso University, Valparaiso, Indiana 46383, USA}
\affiliation{Variable Energy Cyclotron Centre, Kolkata 700064, India}
\affiliation{Warsaw University of Technology, Warsaw, Poland}
\affiliation{University of Washington, Seattle, Washington 98195, USA}
\affiliation{Wayne State University, Detroit, Michigan 48201, USA}
\affiliation{Yale University, New Haven, Connecticut 06520, USA}
\affiliation{University of Zagreb, Zagreb, HR-10002, Croatia}

\author{L.~Adamczyk}\affiliation{AGH University of Science and Technology, Cracow, Poland}
\author{J.~K.~Adkins}\affiliation{University of Kentucky, Lexington, Kentucky, 40506-0055, USA}
\author{G.~Agakishiev}\affiliation{Joint Institute for Nuclear Research, Dubna, 141 980, Russia}
\author{M.~M.~Aggarwal}\affiliation{Panjab University, Chandigarh 160014, India}
\author{Z.~Ahammed}\affiliation{Variable Energy Cyclotron Centre, Kolkata 700064, India}
\author{I.~Alekseev}\affiliation{Alikhanov Institute for Theoretical and Experimental Physics, Moscow, Russia}
\author{J.~Alford}\affiliation{Kent State University, Kent, Ohio 44242, USA}
\author{C.~D.~Anson}\affiliation{Ohio State University, Columbus, Ohio 43210, USA}
\author{A.~Aparin}\affiliation{Joint Institute for Nuclear Research, Dubna, 141 980, Russia}
\author{D.~Arkhipkin}\affiliation{Brookhaven National Laboratory, Upton, New York 11973, USA}
% ...

\collaboration{STAR Collaboration}\noaffiliation
\begin{abstract}

We report the first measurement of charmed-hadron (\dzero ) production via the hadronic decay channel ($D^0\rightarrow K^- + \pi^+$) in \AuAu\ collisions at \sqrtsNN = 200\,GeV with the STAR experiment. The charm production cross-section per nucleon-nucleon collision at mid-rapidity scales with the number of binary collisions, \nbin , from \pp\ to central \AuAu\ collisions. The \dzero\ meson yields in central \AuAu\ collisions are strongly suppressed compared to those in \pp\ scaled by \nbin ,  for transverse momenta $p_{T}>3$ \gevc , demonstrating significant energy loss of charm quarks in the hot and dense medium. 
An enhancement at intermediate \pt\ is also observed.
Model calculations including strong charm-medium interactions and coalescence hadronization describe our measurements.

\end{abstract}

\maketitle


Experimental results from the Relativistic Heavy Ion Collider (RHIC) and the Large Hadron Collider (LHC) support the hypothesis that a strongly-coupled nuclear medium
with partonic degrees of freedom, namely the Quark-Gluon Plasma (QGP), is created in heavy-ion collisions at high energy~\cite{STARwhitepaper}. This state of deconfined matter is of interest to study the nature of the strong force in the unique environment of nuclear matter under extreme energy density.
Charm and beauty quarks are created predominantly via initial hard scatterings in nucleon-nucleon collisions and the production rate is calculable with perturbative QCD techniques~\cite{lin,cacciari}.  The large masses are expected to be retained during the interactions with the nuclear medium. Heavy quarks are therefore predicted to be sensitive to
transport and other properties of the early stages of the system when the QGP is expected to exist~\cite{HFmass}. 

Energetic heavy quarks were predicted to lose less energy via gluon radiation than light quarks when they traverse the QGP~\cite{dead}. Initial RHIC and LHC measurements, however,
show similar suppression at high transverse momentum, \pt , in central \AA\ collisions~\cite{starcharmraa,phenixcharmraa,LHCHF}. This has led to the reconsideration of the effects of heavy-quark collisional energy loss~\cite{Magdelena,Ivancoll} and requires follow-up measurements.

Heavy-quark collective motion can provide experimental evidence for bulk medium thermalization~\cite{teaney}.
Model calculations show that interactions between heavy quarks and the QGP are sensitive to the drag/diffusion coefficients of the medium. These can be related to the shear-viscosity-to-entropy ratio and other transport properties~\cite{teaney}.
Therefore, measurements of heavy-quark production at low and intermediate \pt\ are of particular relevance to these issues, and also for the interpretation of the charmonia production in heavy-ion collisions.

In elementary collisions, heavy quarks are expected to hadronize mainly through hard fragmentation. 
In high-energy heavy-ion collisions, the large charm-pair abundance could increase the coalescence probability.
The coalescence of charm with a  light quark from the medium with a large radial flow may introduce a $p_T$-dependent modification to the observed charmed hadron spectrum compared to that from fragmentation~\cite{TAMU,SUBATECH}. Furthermore, this may lead to a baryon-to-meson enhancement for charmed hadrons similar to that observed for light-flavor hadrons~\cite{B2M,LFraa}. 

In this letter, we report the first measurement of \dzero\ ($c\bar{u}$) production over a transverse momentum range, 0.0 $<p_{\rm T}\lesssim$ 6.0 GeV/$c$, in \AuAu\ collisions at a center of mass energy \sqrtsNN\ of 200 GeV.
The measurement was performed via invariant-mass reconstruction of the hadronic decay channel, $D^0\rightarrow K^-+\pi^+$ and its charge conjugate.
The data used for this analysis were recorded with the Solenoidal Tracker at RHIC (STAR) experiment~\cite{star} during the 2010 and 2011 runs. 
A total of $\sim$8.2$\times 10^8$ minimum-bias-triggered (MB) events and $\sim$2.4$\times 10^8$ 0$-$10\% most central events are used. 
The MB trigger condition is defined as a coincidence signal between the east and west Vertex Position Detectors (VPD)~\cite{vpd} located at $4.4<|\eta|<4.9$. In this analysis, 0$-$80\% of the total hadronic cross section is selected and the $\sim$12\% VPD triggering inefficiency, mostly in the peripheral collisions, is corrected using a Monte Carlo (MC) Glauber simulation~\cite{starsys}.
The most central events (0$-$10\%, corresponding to an average impact-parameter $\sim$3.2 fm) are selected with a combination cut using the spectator signals in the Zero Degree Calorimeter (ZDC)~\cite{zdc} and the multiplicity in the Time-Of-Flight detector (TOF)~\cite{tof} at mid-rapidity. The main subsystems used for the \dzero\ analysis are the Time Projection Chamber (TPC)~\cite{tpc} and the TOF. All measurements are presented as an average of \dzero\ and $\overline{D^0}$ yields at mid-rapidity ($|y|<1$).

In this analysis, the collision vertex position along the beam axis, $V_{\rm z}^{\rm TPC}$, as reconstructed from tracks in the TPC is selected to be within $\pm$30 cm of the nominal center of the STAR detector. In addition, to reject pile-up, the distance between $V_{\rm z}^{\rm TPC}$ and $V_{\rm z}^{\rm VPD}$ obtained from the VPD is required to be less than 3 cm.
The analysis techniques are identical to those for \dAu\ and \pp\ data~\cite{dAuCharm, ppCharm}. 
A cut on the distance of closest approach to the collision vertex of less than 2 cm is required for tracks.
Tracks are required to have at least 20 hits (out of a possible total of 45) to ensure good momentum resolution, and more than 10 hits in the calculation of the ionization energy loss, $\la dE/dx \ra$, to ensure good resolution for particle identification. 
To ensure tracks are reconstructed within the TPC acceptance \pt\ $>$ 0.2 \gevc\ and $|\eta| < 1$ are required. Pions and kaons are separated well by TOF up to $p_T =$ 1.6 \gevc, elsewhere only TPC information is used. At low \pt\ the kaon and pion candidates are identified by combining \dedx\ with timing information with similar cuts on normalized \dedx\ and particle velocity $\beta$ as in the \pp\ analyses~\cite{ppCharm}. However, a tighter kaon identification is used to reduce combinatorial background, specifically \dedx\ is required to be within $\pm$2$\sigma$ of the expected value from Bichsel function calculations~\cite{Bichsel}. 

\begin{figure}[tb]
\includegraphics[width=0.5\textwidth]{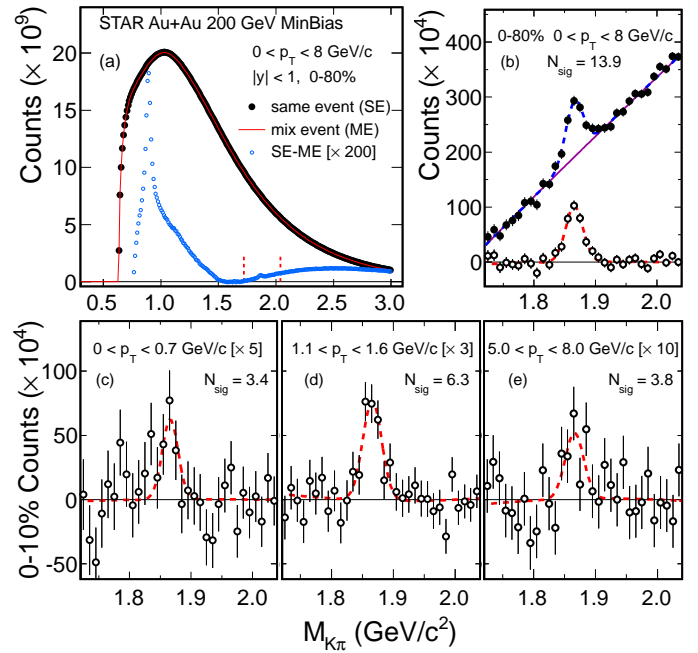}\vspace{-0.25cm}
\caption{(Color online) Panel (a): Invariant mass distribution for all the combinations of kaon and pion candidates  (solid circles). The ME technique reproduces the combinatorial background as shown by the curve. The distribution after ME background subtraction is shown as open circles. Panel (b): $M_{K\pi}$ distribution after ME background subtraction (solid circles) and after further residual background subtraction (open circles). Panels (c)(d)(e) are $M_{K\pi}$ distributions for the 0$-$10\% most central collisions in three \pt\ regions.}
\vspace{-0.35cm}\label{fig:figure1}
\end{figure}

The invariant mass of kaon and pion pairs, $M_{K\pi}$, is constructed from all same-event pair combinations. 
To estimate the combinatorial background from random combinations, the distribution was evaluated using kaon and pion tracks from different collision events with similar characteristics, the mixed-event (ME) technique~\cite{MEtech}.
The $M_{K\pi}$ distribution for 0$-$80\% MB collisions in the range of $0 < p_{T} < 8$ \gevc\ is shown as the solid circles in Fig.~\ref{fig:figure1} (a). The red curve shows that the ME distribution reproduces the combinatorial background. The ME technique introduces negligible ($<$1\%) uncertainties in the \dzero\ signal yields. The open circles represent the $M_{K\pi}$ distribution after the ME background subtraction (scaled by a factor of 200 for visualization). A significant $K$*(892) peak is clearly seen and the \dzero\ signal around 1.86 \gevcc\ is also visible at this scale. 
The solid circles in Fig.~\ref{fig:figure1} (b) show the $M_{K\pi}$ distribution after ME background subtraction in the mass range between the vertical dashed lines in Fig.~\ref{fig:figure1} (a). A quadratic polynomial is used, together with a Gaussian distribution to capture the signal, to fit and subtract residual correlated background. The result is shown as open circles.
The significance, $N_{sig}$, of this signal, calculated as the ratio of the raw yield and the statistical uncertainty including the propagated uncertainties from background subtraction, is 13.9. The mean value of the Gaussian is 1866 $\pm$ 1 MeV, which can be compared to the PDG value (1864.83 $\pm$ 0.14 MeV)~\cite{pdg}. The width (14 $\pm$ 1 MeV) is driven by the detector resolution and is consistent with previous measurements~\cite{ppCharm} and simulations. The mass is constrained to the PDG value for all centrality and \pt\ bins in subsequent fits. Figures~\ref{fig:figure1} (c)(d)(e) show the $M_{K\pi}$ distributions for the 0$-$10\% most central collisions at low \pt , 0$-$0.7 \gevc , intermediate \pt , 1.1$-$1.6 \gevc\ and high \pt , 5.0$-$8.0 \gevc , respectively. 
The significances of the three signals are 3.4, 6.3, and 3.8, respectively. The \dzero\ raw yields are the average values from the fits and from event counting in a $\pm 3 \sigma$ window around the \dzero\ mass. The systematic uncertainties include their differences. The effects of double counting due to particle misidentification have been corrected using the method in Ref.~\cite{ppCharm}.

The raw signal is corrected for the detector acceptance and efficiencies, which are decomposed as the TPC tracking efficiency, the TOF matching efficiency and the particle identification efficiency. The run conditions were similar in 2010 and 2011. A slight difference of the detector performance is reflected in the single track efficiencies. 
This is estimated by first calculating the single pion and kaon track efficiencies via the STAR standard embedding procedure. A number of pions or kaons equal to 5\% of the event's multiplicity are simulated through the STAR detector geometry in GEANT and embedded into the real event, followed by the standard offline reconstruction. The single track efficiency is calculated by comparing the reconstructed tracks with the MC input tracks. 
The track efficiency includes the net effect from track splitting and merging, TPC acceptance, decays and interaction losses in the detector.
The TOF matching and particle identification efficiencies are evaluated based on the distributions in the data. The \dzero\ efficiency is calculated via the single track efficiencies in each \pt, $\eta$ and $\phi$ bin by folding with the decay kinematics.

The systematic uncertainties in the \pt\ spectra include: $\rm a$) \dzero\ raw yield extraction uncertainties, 1\% at 2\,GeV/$c$ then increasing to 9\%(10\%) at the lowest (highest) \pt\ bin, $\rm b$) efficiency uncertainties, 11\% at low \pt\ then slowly decreasing to 9\% at high \pt\, $\rm c$) overall charm fragmentation ratio uncertainty, 5.7\% and \dzero\ decay branching ratio uncertainty, 1.3\%. When calculating the \dzero\ nuclear modification factor (\RAA) which will be described later, uncertainties in ($\rm c$) are cancelled and the efficiency uncertainties in ($\rm b$) are largely reduced because of the same detector system. 
However, the following additional uncertainties contribute to the \dzero\ $R_{\rm AA}$: $\rm d$) uncertainties of the \pt\ spectrum in \pp\ collisions including the functional extrapolation to unmeasured \pt\, 10\% at 2\,GeV/$c$ then increasing to 35\%(30\%) at the lowest (highest) \pt\ bin, $\rm e$) overall uncertainties of \nbin\ in different centralities, which are listed in Table~\ref{Tglauber}. 

\begin{figure}[tb]
\includegraphics[width=0.45\textwidth]{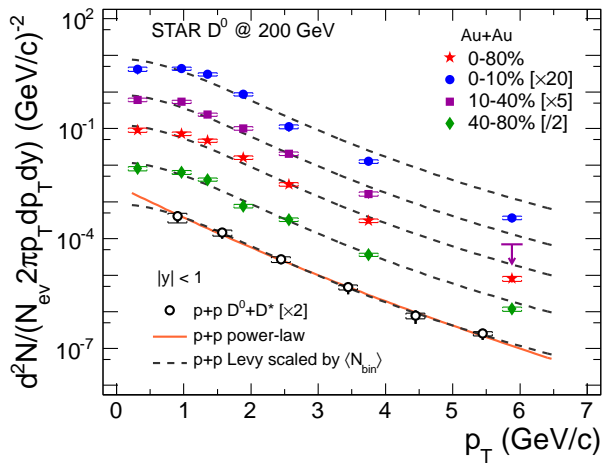}\vspace{-0.25cm}
\caption{(Color online) Centrality dependence of the \dzero\ \pt\ differential invariant yield in \AuAu\ collisions (solid symbols). The curves are number-of-binary-collision-scaled Levy functions from fitting to the \pp\ result (open circles)~\cite{ppCharm}. The arrow denotes the upper limit with 90\% confidence level of the last data point for 10$-$40\% collisions. The systematic uncertainties are shown as square brackets.} \vspace{-0.35cm}\label{fig:figure2}
\end{figure}


The \dzero\ \pt\ spectra after corrections are shown in Fig.~\ref{fig:figure2} as solid symbols for different centrality bins. The $D^0$ and charm production cross sections are extracted from the integration of the \dzero\ \pt\ spectra and the uncertainties are obtained following the method used in Ref.~\cite{ppCharm}. The $D^0$ per nucleon-nucleon-collision production cross section, $d\sigma_{D\bar{D}}^{NN}/dy$, in the 0$-$10\% most central collisions is measured to be 84 $\pm$ 9 (stat.) $\pm$ 10 (sys.) $\mu$b. The charm $d\sigma_{c\bar{c}}^{NN}/dy$ at mid-rapidity in the 0$-$10\% most central collisions is calculated, assuming the same $c \rightarrow D^{0}$ fragmentation ratio (0.565 $\pm$ 0.032) as in \pp\ collisions~\cite{ppCharm}, to be 148 $\pm$ 15 (stat.) $\pm$ 19 (sys.) $\mu$b.  

\begin{figure}[tb]
\includegraphics[width=0.45\textwidth]{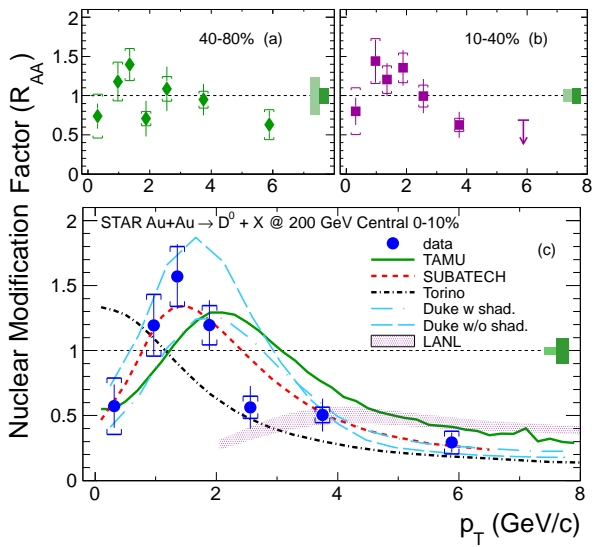}\vspace{-0.25cm}
\caption{(Color online) Panels (a)(b): \dzero\ \RAA\ for peripheral 40$-$80\% and semi-central 10$-$40\% collisions; Panel (c): \dzero\ \RAA\ for 0$-$10\% most central events (blue circles) compared with model calculations from the TAMU (solid curve), SUBATECH (dashed curve), Torino (dot-dashed curve), Duke (long-dashed and long-dot-dashed curves), and LANL groups (filled band). The vertical lines and boxes around the data points denote the statistical and systematic uncertainties. The vertical bars around unity denote the overall normalization uncertainties in the \AuAu\ and \pp\ data, respectively.}\vspace{-0.35cm}\label{fig:figure3}
\end{figure}

The \pp\ data, shown as open circles, contain \dzero\ data for \pt\ $<$ 2.0 \gevc\ and $D^{*}$ data for \pt\ $>$ 2.0 \gevc~\cite{ppCharm}. The dashed curves are Levy function~\cite{Levy} fits to the \pp\ data, scaled by the number of binary collisions, \nbin ~\cite{starsys}. Table~\ref{Tglauber} contains the values of \nbin\ and of \npart , the number of participants. 
The \dzero\ \RAA\ is calculated as the ratio between the \dzero\ \pt\ spectrum in \AuAu\ collisions in each centrality bin to the Levy function fit
to the \pp\ data scaled by \nbin ~\cite{ppCharm}. 
The difference between power-law and Levy functions is taken into account in the bin-by-bin systematic uncertainties; especially for the low \pt\ extrapolation where the data points are missing in the \pp\ data. Figure~\ref{fig:figure3} shows \dzero\ \RAA\ for the centrality bins of 40$-$80\% (a), 10$-$40\% (b) and 0$-$10\% (c). The vertical lines and brackets indicate the size of the statistical and systematic uncertainties, respectively. The vertical bars around unity from left to right represent the overall scaling uncertainties for \nbin\ in \AuAu\ and cross-section in \pp\ collisions, respectively. Strong suppression is observed in the most central collisions for \pt\ $>$ 2.5 \gevc , while no evidence is found for suppression in peripheral collisions. In 0$-$10\% collisions, the suppression level is around 0.5 for \pt\ $>$ 3 \gevc , which is consistent with both the measurements of electrons from heavy flavor hadron decays~\cite{starcharmraa, phenixcharmraa} and the light hadrons~\cite{LFraa}. 
This indicates that charm quarks lose energy as they pass through the medium in central \AuAu\ collisions at RHIC. In Pb+Pb collisions at 2.76 TeV at the LHC, a strong suppression of leptons and $D$-mesons~\cite{LHCHF} has also been observed equal to that of the light hadrons.

\begin{table}
\begin{center}
\vskip 0.1 in \centering\begin{tabular}{lr@{.}lr@{.}lr@{.}lr@{.}lr@{.}lr@{.}lr@{.}lr@{.}l} \hline \hline
centrality & \multicolumn{4}{c}{0$-$10\%} & \multicolumn{4}{c}{10$-$40\%} & \multicolumn{4}{c}{40$-$80\%} & \multicolumn{4}{c}{0$-$80\%} \\
\hline
$N_{\mathrm{bin}}$  & 941&2 $\pm$& 26&3 & 391&4 $\pm$& 30&2 & 56&6 $\pm$& 13&6 & 291&9 $\pm$& 20&5 \\
$N_{\mathrm{part}}$ & 325&5 $\pm$& 3&6 & 174&1 $\pm$& 9&9 & 41&8 $\pm$& 7&8 & 126&7 $\pm$& 7&7 \\
\hline \hline
\end{tabular}
\vskip 0.1 in
\caption{The number of binary collisions and the number of participants from Glauber model calculation~\cite{starsys}.} 
\label{Tglauber}
\end{center}
\end{table}

Several recent model calculations are compared with STAR data in Fig.~\ref{fig:figure3} (c).  
The TAMU group~\cite{TAMU} used the Langevin approach to calculate heavy quark propagation in the medium described by a (2+1)D ideal hydrodynamic model. 
The charm-medium interaction strength is calculated using a T-Matrix dynamic method. 
The calculation considered collisional energy loss and charm-quark hadronization including both fragmentation and coalescence mechanisms.
The SUBATECH group~\cite{SUBATECH} used the Hard-Thermal-Loop (HTL) analytic approach to calculate charm-medium interactions with both fragmentation and coalescence processes. 
Their calculations suggest
that the radiative energy loss has a negligible impact on the final charmed hadron \RAA\ for $p_{\rm T}<6$\, GeV/$c$.
The Torino group~\cite{Torino} applied the HTL calculation of the charm-medium interaction strength into the Langevin simulation with the medium described via viscous hydrodynamics. 
However, this calculation does not include the charm-quark coalescence hadronization process.
The calculations from the TAMU and SUBATECH groups generally describe the significant features in the data, while the Torino calculation misses the intermediate-\pt\ enhancement structure with a $\chi^{2}=$ 16.1 for five degrees of freedom for $p_{T}<3$ \gevc , considering the quadratic sum of statistical and systematic uncertainties.
This indicates that, in the measured kinematic region, collisional energy loss alone can account for the large suppression in \RAA , but a coalescence type mechanism is important in modeling charm-quark hadronization at low and intermediate \pt.
Cold-nuclear-matter (CNM) effects in the open charm sector could also be important, and could contribute to the enhancement of \RAA . Calculations from the Duke group~\cite{Duke}, including fragmentation and recombination with or without shadowing effects provide a reasonable description of the data.  The treatment from the LANL group~\cite{CNM} with CNM and hot
QGP effects, including energy loss and meson dissociation, is consistent in the region of its applicability, $p_T > 2$ \gevc , with our data. 
At LHC energies, all these models reproduce the strong suppression of D-meson production in central Pb+Pb collisions at $p_{T} >$ 2 \gevc . However, no data is available from LHC to justify these models at $p_{T} <$ 2 \gevc ~\cite{LHCHF}.

\begin{figure}[tb]
\includegraphics[width=0.4\textwidth]{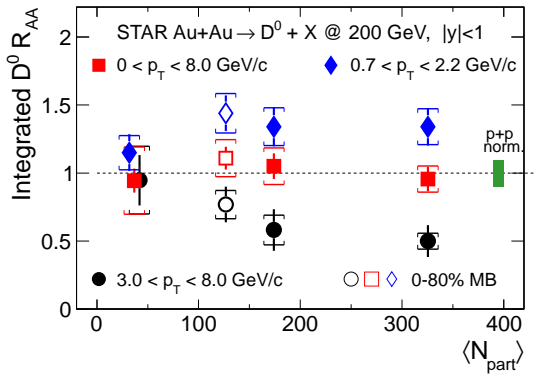}\vspace{-0.25cm}
\caption{(Color online) Integrated \dzero\ \RAA\  as a function of \npart in different \pt\ regions: 0$-$8 \gevc\ (square), 3$-$8 \gevc\ (circles) and 0.7$-$2.2 \gevc\ (diamonds). 
Open symbols are for the 0$-$80\% MB events. 
} \vspace{-0.35cm}\label{fig:figure4}
\end{figure}

The integrated \RAA\ is calculated as the ratio of the integrated yield in Au+Au collisions divided by the integration of the \pp\ reference as above scaled by the number of binary collisions in the given \pt\ region. Figure~\ref{fig:figure4} shows the integrated \dzero\ \RAA\ as a function of \npart. The red squares represent the integrated \RAA\ over the whole \pt\ region, which agree with unity, indicating that the charm production cross section scales with the number of binary collisions. This is consistent with charm quarks originating predominantly from initial hard scattering at RHIC. The integrated \RAA\ above 3 \gevc\ are represented as black circles, and show a strong centrality dependence. No suppression is seen in peripheral collisions, but a clear suppression, at the level of $\sim$0.5, is seen in central collisions. An enhancement is observed from the \RAA\ integrated over the intermediate \pt\ region 0.7$-$2.2 \gevc , shown as blue diamonds.


In summary, we report the first \dzero\ production measurement via $D^0\rightarrow K^- + \pi^+$ decay at mid-rapidity in \sqrtsNN = 200\,GeV \AuAu\ collisions.
The charm production cross sections at mid-rapidity per nucleon-nucleon collision from \pp\ to \AuAu\ show a number-of-binary-collision scaling, which supports that charm quarks are mainly produced in the initial hard scatterings. 
The centrality dependence of the \pt\ distributions as well as the nuclear modification factor show no suppression in peripheral collisions, but a strong suppression, at the level of \RAA\ $\sim0.5$, in the most central collisions for \pt\ $>$ 3 \gevc .
This is indicative of significant energy loss of charm quarks in the hot dense medium. 
An enhancement in the intermediate-\pt\ region is also observed for the first time in heavy-ion collisions for charmed mesons.  The $D^0$ \RAA\ is consistent with model calculations including strong charm-medium interactions and hadronization via coalescence at intermediate \pt . 


We thank the RHIC Operations Group and RCF at BNL, the NERSC Center at LBNL, the KISTI Center in Korea, and the Open Science Grid consortium for providing resources and support. This work was supported in part by the Offices of NP and HEP within the U.S. DOE Office of Science, the U.S. NSF, CNRS/IN2P3, FAPESP CNPq of Brazil,  the Ministry of Education and Science of the Russian Federation, NNSFC, CAS, MoST and MoE of China, the Korean Research Foundation, GA and MSMT of the Czech Republic, FIAS of Germany, DAE, DST, and CSIR of India, the National Science Centre of Poland, National Research Foundation (NRF-2012004024), the Ministry of Science, Education and Sports of the Republic of Croatia, and RosAtom of Russia.

\end{document}